\documentclass[aps,prl,twocolumn,superscriptaddress,showpacs,floatfix]{revtex4}
\usepackage{graphicx}
\usepackage{epsfig}
\usepackage{amsmath}
\usepackage{amssymb}

\begin{document}

\title{Structure and properties of an amorphous metal--organic framework}

\author{Thomas D. Bennett}
\affiliation{Department of Materials Science, University of Cambridge, Cambridge CB2 3QZ, U.K.}

\author{Andrew L. Goodwin}
\affiliation{Department of Earth Sciences, Cambridge University,
Downing Street, Cambridge CB2 3EQ, U.K.}
\affiliation{Department of Chemistry, University of Oxford, South Parks Road, Oxford OX1 3QR, U.K.}

\author{Martin T. Dove}
\affiliation{Department of Earth Sciences, Cambridge University,
Downing Street, Cambridge CB2 3EQ, U.K.}

\author{David A. Keen}
\affiliation{ISIS Facility, Rutherford Appleton Laboratory, Harwell Science and Innovation Campus, Didcot, Oxfordshire OX11 0QX, U.K.}
\affiliation{Department of Physics, Oxford University, Clarendon Laboratory, Parks Road, Oxford OX1 3PU, U.K.}

\author{Matthew G. Tucker}
\affiliation{ISIS Facility, Rutherford Appleton Laboratory, Harwell Science and Innovation Campus, Didcot, Oxfordshire OX11 0QX, U.K.}

\author{Emma R. Barney}
\affiliation{ISIS Facility, Rutherford Appleton Laboratory, Harwell Science and Innovation Campus, Didcot, Oxfordshire OX11 0QX, U.K.}

\author{Alan K. Soper}
\affiliation{ISIS Facility, Rutherford Appleton Laboratory, Harwell Science and Innovation Campus, Didcot, Oxfordshire OX11 0QX, U.K.}

\author{Erica G. Bithell}
\affiliation{Department of Materials Science, University of Cambridge, Cambridge CB2 3QZ, U.K.}

\author{Jin-Chong Tan}
\affiliation{Department of Materials Science, University of Cambridge, Cambridge CB2 3QZ, U.K.}

\author{Anthony K. Cheetham}
\email{akc30@cam.ac.uk}
\affiliation{Department of Materials Science, University of Cambridge, Cambridge CB2 3QZ, U.K.}

\date{\today}
\begin{abstract}
We show that ZIF-4, a metal-organic framework (MOF) with a zeolitic structure, undergoes a crystal--amorphous transition on heating to 300\,$^\circ$C. The amorphous form, which we term a-ZIF, is recoverable to ambient conditions or may be converted to a dense crystalline phase of the same composition by heating to 400\,$^\circ$C. Neutron and X-ray total scattering data collected during the amorphization process are used as a basis for reverse Monte Carlo refinement of an atomistic model of the structure of a-ZIF. We show that the structure is best understood in terms of a continuous random network analogous to that of a-SiO$_2$. Optical microscopy, electron diffraction and nanoindentation measurements reveal a-ZIF to be an isotropic glass-like phase capable of plastic flow on its formation. Our results suggest an avenue for designing broad new families of amorphous and glass-like materials that exploit the chemical and structural diversity of MOFs.
\end{abstract}


\maketitle

\begin{figure}
\begin{center}
\includegraphics{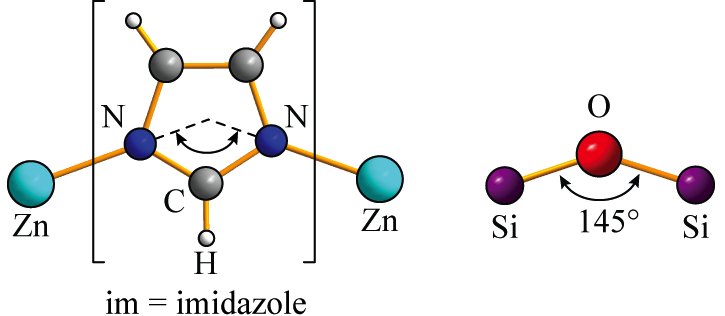}
\end{center}
\caption{\label{fig1}Representations of the Zn--im--Zn and Si--O--Si linkages in tetrahedral ZIF and silicate networks, respectively.}
\end{figure}

Amorphous materials are found extensively in the fields of intermetallic \cite{Greer_1995}, inorganic \cite{Greaves_2007}, and organic materials \cite{Saragi_2007}, and have a wide range of important applications. Throughout the last 5--10 years the materials science community has studied extensively a fourth class of materials: namely, the so-called metal-organic frameworks (MOFs) \cite{Ferey_2009,Rao_2008}. These are network structures in which metal atoms are combined with organic ligands in a crystalline array. Their structures can be dense or porous and, by virtue of their combined structural and chemical versatility, they have a wide range of applications, including gas sorption and separation, magnetism and photoluminescence. While amorphous products have been identified in this emerging area \cite{Oh_2005,Maspoch_2003,Masciocchi_2003} (including a very recent report of pressure-induced amorphization \cite{Chapman_2009}) there remain no well-characterized examples. The zeolitic imidazolate frameworks (ZIFs) \cite{Lehnert_1980,Tian_2003,Banerjee_2008} are interesting candidates for amorphization because their structures are closely related to those of certain zeolitic silica polymorphs by virtue of their analogous tetrahedral connectivites [Fig.~\ref{fig1}]. Just as silica-based materials can form network glasses, we show here that ZIFs can also be prepared in silica-glass-like forms. Our results demonstrate how the chemical versatility of MOFs might be coupled with the attractive mechanical, optical and electronic properties of the amorphous state to yield a new generation of advanced functional materials.

Inorganic zeolites can transform to amorphous structures under the application of either pressure or temperature \cite{Greaves_2003}. In the present work, we have discovered that the zinc imidazolate ZIF-4 \cite{Park_2006} will amorphize at ambient pressure, simply on heating. Bulk samples of ZIF-4 were prepared via solvothermal reaction of zinc(II) nitrate hexahydrate and imidazole in dimethylformamide (DMF) \cite{Park_2006}. Thermogravimetric analysis and variable-temperature X-ray and neutron diffraction data all show that the as-prepared phase, which contains DMF molecules within its framework pore structure, loses this DMF on heating above 200\,$^\circ$C, giving a solvent-free nanoporous phase of composition Zn(im)$_2$ (im$^-$ = imidazolate anion) while maintaining the ZIF-4 topology. Further heating to 300\,$^\circ$C causes an irreversible transformation to an amorphous phase (referred to hereafter as a-ZIF), indicated by the loss of all Bragg reflections in the diffraction pattern. This phase is recoverable to ambient temperature with no change in its diffraction pattern. On further heating to 400\,$^\circ$C the densest of the known crystalline Zn(im)$_2$ phases (ZIF-zni) \cite{Lehnert_1980} evolves from the amorphous solid and is stable to around 500\,$^\circ$C, whereupon thermal decomposition takes place. The overall transformation of ZIF-4 to ZIF-zni is consistent with the greater thermodynamic stability of the latter, predicted by recent density functional theory calculations \cite{Lewis_2009}. While the existence of other amorphous transition-metal imidazolates has been documented \cite{Masciocchi_2001,Masciocchi_2001b,Masciocchi_2003,Chapman_2009}, the unique advantage of having discovered this particular system is that the thermal stability field of a-ZIF is bounded by crystalline phases of the same composition Zn(im)$_2$, removing any ambiguities of composition during structure determination.

\begin{figure}
\begin{center}
\includegraphics{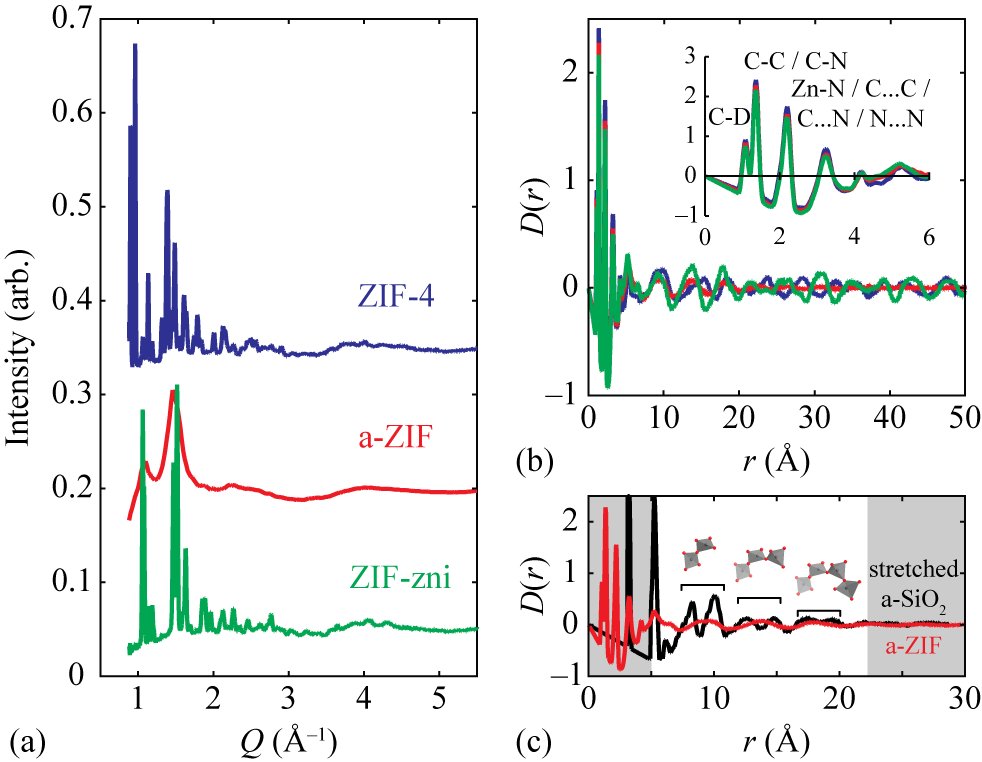}
\end{center}
\caption{\label{fig2}(a) Low-angle ($Q<5.5$\,\AA$^{-1}$) region of the neutron total scattering patterns of (top) the desolvated ZIF-4 framework at 300\,$^\circ$C, (middle) the \emph{in-situ} amorphized a-ZIF product at 320\,$^\circ$C, and (bottom) subsequently-recrystallised ZIF-zni at 400\,$^\circ$C. (b) Neutron differential correlation functions $D(r)$ \cite{Keen_2001}; the inset illustrates the practically-identical local structure in all phases. (c) $D(r)$ functions compared for a-ZIF and  a-SiO$_2$ \cite{Tucker_2005}, where the latter data have been stretched in $r$ by a factor of two to reflect approximately the difference in Zn$\ldots$Zn and Si$\ldots$Si distances in the two materials.}
\end{figure}

In order to probe the detailed structure of a-ZIF, we prepared a perdeuterated sample of ZIF-4 and collected neutron total scattering data at a number of temperatures using the GEM diffractometer at the ISIS spallation source \cite{Hannon_2005}. Neutron powder diffraction patterns of the desolvated (300\,$^\circ$C) ZIF-4 phase, the as-prepared amorphous phase (320\,$^\circ$C), and the recrystallized ZIF-zni phase (400\,$^\circ$C) are shown in Fig.~\ref{fig2}(a). Following appropriate normalization (see SI), the data were converted to the corresponding pair distribution functions (PDFs) [Fig.~\ref{fig2}(b)]. Further structural data were collected for the a-ZIF and ZIF-zni phases using Ag K$\alpha$ X-ray radiation ($\lambda=0.561$\,\AA) \cite{tenijenhuis_2009}; the complete set of X-ray total scattering patterns and their corresponding Fourier transforms (\emph{i.e.} effective PDFs \cite{Keen_2001}) are given as SI.

The most striking feature of the PDFs --- both neutron and X-ray --- is that they are virtually identical for each of ZIF-4, a-ZIF and ZIF-zni up to approximately 6\,\AA\ (see inset to Fig.~1(b) and SI), confirming that the tetrahedral Zn coordination environment and the bridging coordination motif of the imidazolate ions [Fig.~\ref{fig1}] are common to all three phases. We note that there is some slight broadening of the peaks in the PDFs with increasing temperature as is consistent with increased thermal motion. The PDFs begin to differ at larger distances with evidence for different medium-range order (MRO) in all three phases over distances of 10--20\,\AA.  Peaks characteristic of crystalline order persist in the PDFs of ZIF-4 and ZIF-zni for $r>20$\,\AA, whereas the PDF of a-ZIF is essentially featureless after $r\simeq25$\,\AA. A very similar comparison can be drawn between crystalline silica polymorphs and amorphous silica (a-SiO$_2$) \cite{Keen_1999}, where the lower critical lengthscales reflect the shorter distance between tetrahedral centres ($d_{\rm Si\ldots Si}\simeq3.1$\,\AA; \emph{cf}.\ $d_{\rm Zn\ldots Zn}\simeq5.9$\,\AA\ for Zn--im--Zn). Indeed there are strong qualitative similarities in the PDFs of a-SiO$_2$ and a-ZIF at longer distances [Fig.~\ref{fig2}(c)] that add further weight to the analogy often drawn between ZIFs and silicate analogues \cite{Park_2006}.

Detailed insight into the structure of a-ZIF was obtained via Reverse Monte Carlo (RMC) refinement against both neutron and X-ray total scattering data using the program RMCProfile \cite{Tucker_2007}. RMC modelling of crystalline--amorphous transitions in zeolites \cite{Haines_2009} and the open framework material ZrW$_2$O$_8$ \cite{Keen_2007} has shown framework connectivity is commonly preserved during framework amorphization. Consequently our initial starting models were based on the crystalline topologies of either ZIF-4 or ZIF-zni frameworks. We found RMC refinements that used these models were not capable of reproducing both neutron and X-ray scattering data, even after incorporation of substantial disorder of the imidazolate linkages during refinement (see SI for further details). The key implication of this result is that there must be a series of changes in connectivity within the Zn(im)$_2$ framework during both the conversion from ZIF-4 to a-ZIF and then again from a-ZIF to ZIF-zni. Indeed this is consistent with the changes in MRO evident in the PDF data themselves [Fig.~\ref{fig2}(b)] and also that ZIFs are known to undergo reconstructive rather than displacive transitions under application of modest pressures \cite{Spencer_2009,Moggach_2009}.

\begin{figure}
\begin{center}
\includegraphics{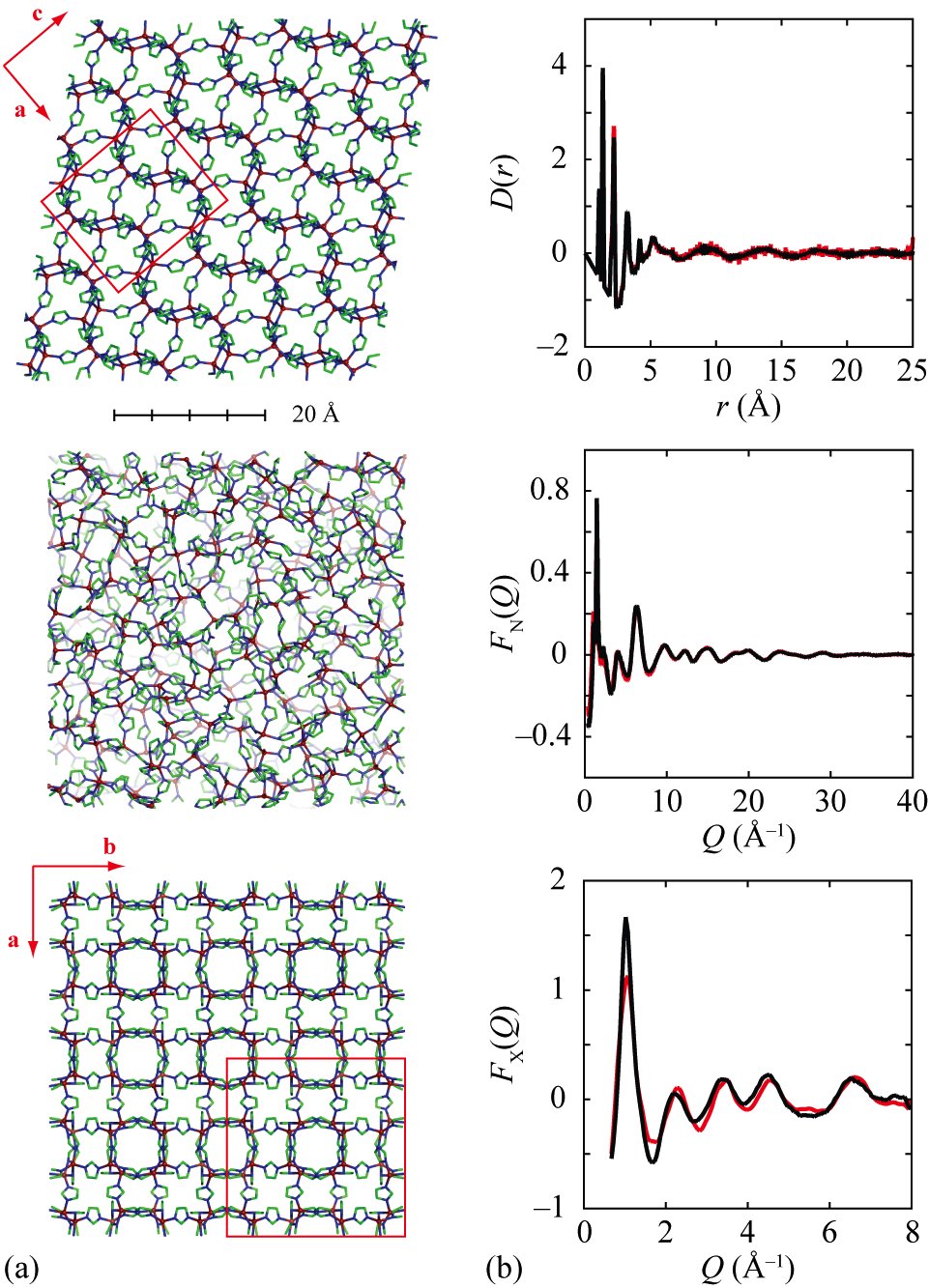}
\end{center}
\caption{\label{fig3}(a) A region of the final RMC model of a-ZIF (centre) shown relative to comparable regions of the ZIF-4 (top) and ZIF-zni (bottom) framework structures; the corresponding unit cells are shown in red outline. Deuterium atoms have been omitted for clarity. (b) RMC fits (red lines) to experimental total scattering data (black lines), calculated using the single atomistic configuration illustrated in the centre panel of (a): (top) Neutron differential correlation function $D(r)$, (centre) neutron total scattering structure factor $F_{\rm N}(Q)$, and (bottom) X-ray total scattering structure function $F_{\rm X}(Q)$ \cite{Keen_2001}.}
\end{figure}

Motivated by the correspondence between features in the a-SiO$_2$ and a-ZIF PDFs shown in Fig.~\ref{fig2}(c), we tested a model based on an established continuous random network (CRN) topology of a-SiO$_2$ itself \cite{Tucker_2005,Wooten_1985}, modified to reflect the replacement of Si by Zn and oxygen by imidazolate. RMC refinement then gave markedly improved fits to data, with the final RMC configuration capable of reproducing all key aspects of the experimental total scattering patterns. The configuration itself is illustrated in Fig.~\ref{fig3}(a), with the corresponding fits shown in Fig.~\ref{fig3}(b). This is strong evidence that the structure of a-ZIF is itself well described by a CRN of tetrahedral Zn centres connected via imidazolate linkages. While the CRN of a-SiO$_2$ probably has different ``ring statistics'' to that of a-ZIF (we note that both ZIF-4 and ZIF-zni networks contain a relatively large number of four-membered rings, while a-SiO$_2$ is thought to contain very few), so far our RMC modelling + total scattering approach is not sensitive to these subtleties. What we can state with confidence is that the network topology in a-ZIF is disordered rather than crystal-like: \emph{the material is ``truly'' amorphous with a CRN topology, and not a highly disordered network solid with a crystal-like topology}.

\begin{figure*}
\begin{center}
\includegraphics{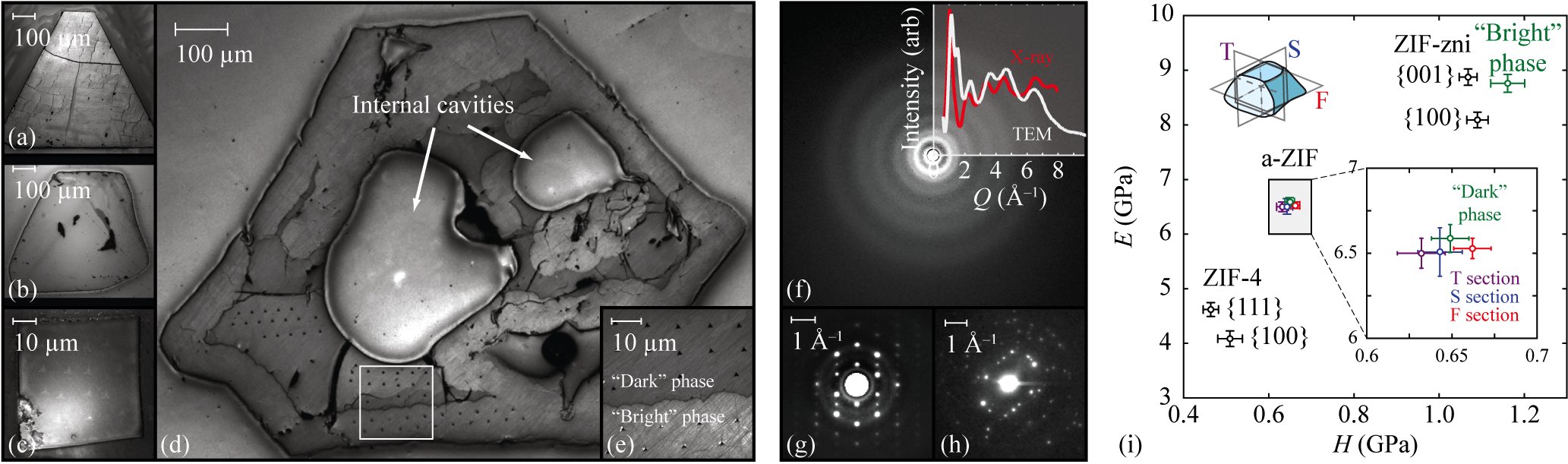}
\end{center}
\caption{\label{fig4}Typical optical micrographs of: (a) a desolvated ZIF-4 single crystal, where the desolvation process has induced visible surface cracking; (b) a recovered a-ZIF monolith, showing curved external surfaces and curved internal cavities; (c) a ZIF-zni single crystal prepared independently (see SI for further details); (d, e) a partially-recrystallized monolith consisting of a-ZIF (``dark'') and ZIF-zni (``bright'') phases --- panel (e) is that portion of the monolith in (d) enclosed within the white square and shows the indentations used to measure mechanical behaviour of the two phases \cite{Tan_2009}. Electron diffraction patterns from: (f) an a-ZIF monolith together with the radially-averaged intensity distribution for qualitative comparison with the X-ray scattering function $F_{\rm X}(Q)$; (g) a crystallite of ZIF-4 prior to heating; (h) a polycrystalline monolith after transformation to ZIF-zni. (i) Young's moduli ($E$) and hardnesses ($H$) determined for the different phases in panels (a)--(e) along known orientations, over surface penetration depths of 150 to 1000\,nm (``T'', ``S'', ``F'' refer to the transverse, sagittal and frontal sections). These measurements were used to assign the ``dark'' and ``bright'' phases in panel (c) as a-ZIF and ZIF-zni, respectively, indicating that ZIF-zni nucleates at the monolith surfaces during recrystallization.}
\end{figure*}

Further evidence for glass-like behaviour in a-ZIF comes from optical microscopy. Crystals of ZIF-4 deform, but do not shatter, on the transition to a-ZIF [Fig.~\ref{fig4}(a,b)]. Quite remarkably the monolithic ``crystals'' evolve curved internal cavities (presumably coupled with loss of included DMF) in addition to curved external morphologies across the transition [Fig.~\ref{fig4}(b,d)], indicating that a-ZIF appears to exhibit viscous flow on its formation at 300\,$^\circ$C. Optical micrographs of partially transformed ``crystals'' show that the subsequent transition from a-ZIF to ZIF-zni then occurs via nucleation at surface defects [Fig.~\ref{fig4}(d,e)]. Electron diffraction patterns collected from the amorphous phase indicate it to be isotropic [Fig.~\ref{fig4}(f)], with no residual texture from the anisotropic ZIF-4 parent phase [Fig.~\ref{fig4}(g)]. Subsequent recrystallisation of ZIF-zni during \emph{in-situ} heating yields polycrystalline monoliths with sub-micron grains of random orientation [Fig.~\ref{fig4}(h)], suggesting there is no ``memory'' of the original ZIF-4 orientation. In light of these findings, one also expects a-ZIF to exhibit mechanical isotropy rather than the anisotropy of the two crystalline phases. We tested this by performing a series of nanoindentation studies \cite{Tan_2009} on single crystals/monoliths of the ZIF-4, a-ZIF and ZIF-zni samples. Elastic moduli and hardness values confirm the loss of anisotropy in the amorphous phase and show the mechanical properties of a-ZIF to be intermediate to but distinct from those of ZIF-zni and ZIF-4 [Fig.~\ref{fig4}(i)].

That a-ZIF is not formed directly from a melt precludes the use of the term ``glass'' in its traditional sense; specifically, there is no glass transition \emph{per se}. Nevertheless it is clear that a-ZIF is fundamentally different from the crystalline ZIF phases. An important question to be answered will be why such a phase should form at all. What we do know is that, like their silicate analogues, ZIFs are only metastable (as indeed are many open-framework MOFs), and this will provide some driving force for thermally-activated reconstructive transitions. The existence of multiple intermediate structure topologies with similar energies will then favour an amorphous state until the crystalline minimum (ZIF-zni) becomes thermally accessible. It appears that ZIF materials follow their silica analogues in this respect, and indeed similar arguments have been put forward to explain the formation of amorphous covalent organic networks \cite{Jiang_2007}.

There is every reason to expect similar amorphization mechanisms in other metal-organic frameworks. Indeed we observe this to be the case in other ZIFs, including those in which Zn is replaced by magnetically-active Co(II). It is precisely this chemical variability that means access to glass-like MOFs will offer a number of exciting opportunities in the development of functional amorphous materials. The use of chiral bridging ligands, for example, suggests a method of producing optically-active network glasses; likewise the incorporation of lanthanide metal centres and/or electronically active ligands would be an obvious means of preparing electroluminescent glasses for advanced photonics. Our results here show that the major hurdle faced by the community in developing such systems --- namely, access to a method of characterising and understanding their atomic-scale structures --- is no longer insurmountable.

We gratefully acknowledge financial support from the European Research Council to A.K.C., from Trinity College, Cambridge to A.L.G., from the E.P.S.R.C. (UK) to A.L.G. and T.D.B., from the Isaac Newton Trust to J.C.T. and from HH Sheikh Saud Bin Saqr Al Qasimi to T.D.B. We thank A.\ Thirumurugan (Cambridge) for useful discussions.



\end{document}